\documentstyle[preprint,aps]{revtex}

\begin{document}

\draft

\preprint{STPHY-Th/95-1}

\title{\bf Quantizing SU(N) gauge theories without gauge fixing}

\author{G B Tupper and F G Scholtz\\[5pt]
\it Institute of Theoretical Physics\\
\it University of Stellenbosch\\
\it 7600 $\;$  Stellenbosch\\
\it South Africa}

\date{January 17, 1995}

\maketitle

\abstract{We generalize and extend the quantization procedure of \cite{st}
which is designed to quantize SU(N) gauge theories in the continuum without
fixing the gauge and thereby avoid the Gribov problem.  In particular we
discuss the BRS symmetry underlying the effective action.  We proceed to use
this BRS symmetry to discuss the perturbative renormalization of the theory and
show that perturbatively the procedure is equivalent to Landau gauge fixing.
This generalizes the result of \cite{st} to the non-abelian case and confirms
the widely held believe that the Gribov problem manifests itself on the
non-perturbative level, while not affecting the perturbative results.  A
relation between the gluon mass and gluon condensate in QCD is obtained which
yields a gluon mass consistent with other estimates for values of the gluon
condensate obtained from QCD sum rules.}


\newpage

\section{Introduction}

A major obstacle in the quantization of non-abelian gauge theories is
the Gribov problem \cite{grib}, especially as formulated by Singer \cite{sing}:
consider a compact, semisimple non-abelian gauge theory in Euclidean
space-time with boundary conditions at infinity implying the identification of
space-time with $S^4$, then due to a topological obstruction no global
continuous gauge fixing is possible.  This excludes a very general class of
gauge fixing conditions and in particular all the practically implementable
ones.

It seems to be generally accepted that the Gribov problem is not important in
the perturbative domain, but that it may play an important role in
understanding the non-perturbative aspects of a gauge theory \cite{crev}.  Our
present results confirm that the Gribov problem is perturbatively unimportant.
However, the situation regarding the non-perturbative aspects is much less
clear. Indeed, Fujikawa \cite{fuj} argued that non-perturbatively the usual BRS
symmetry \cite{brs} is spontaneously broken when a Gribov problem is present,
invalidating the associated Slavnov-Taylor identities.  Thus, while
perturbatively innocuous, the Gribov problem casts doubt on e.g. the program of
solving the Schwinger-Dyson equations non-perturbatively via the gauge
technique \cite{bak}.  It is therefore of paramount importance that the Gribov
problem should be brought under control before reliable investigations into
the non-perturbative aspects of gauge theories can be launched.

Within the continuum limit there has been two recent proposals for quantizing
non-abelian gauge theories without gauge fixing and thereby
avoiding the Gribov problem.  The first of these, 'soft gauge fixing', due to
Zwanziger \cite{zwan} and Jona-Lasinio \cite{jl} amounts to an implementation
of Popov's suggestion \cite{pop} that the Faddeev-Popov trick should be
generalized to $1=\int[dU]F[ ^UG]/I[G]$ with $I[G]=\int[dU]F[ ^UG]$ and $F$ a
non-gauge invariant function of the gauge field $G$ such that $I[G]$ exists.
For a particular choice of $F[G]$ this method has been shown by Fachin
\cite{fach} to reproduce the usual perturbative result for the renormalized
Landau-gauge propogator in a suitable limit.

The second approach is due to the present authors \cite{st}.  Here the point of
departure is a supersymmetry-like resolution of the identity in terms of
bosonic and fermionic auxiliary fields combined with a U-gauge transformation
to give the gauge field a mass.  Since the starting identity has a BRS symmetry
one also expects this to be the case for the effective action resulting from
this quantization scheme. This is indeed the case as is shown below.

There are, however, several outstanding problems connected to the procedure of
\cite{st}.  Firstly it has not yet been demonstrated how the usual
perturbative results can be recovered from this procedure for a non-abelian
theory. Indeed one may appreciate that since the program invokes a massive
U-gauge like non-abelian gauge field the ordinary perturbation theory and
renormalization analysis are problematic.  Secondly it can be shown (see
section 6) that the original proposition of breaking the gauge symmetry at the
tree level is perturbatively incompatible with the BRS invariance as it would
imply the spontaneous breaking of the BRS symmetry. Thirdly
the sequential scheme originally proposed to deal with SU(N$>$2) is clumsy and
obscures the BRS symmetry underlying the quantization procedure.

In this paper we will show how these problems are overcome.  We begin in
section 2 by reformulating our procedure so as to effectively deal with any
SU(N$>$2) theory without resorting to the sequential scheme of \cite{st}.  This
is done by imbedding in a U(N) gauge theory with the identity realized in terms
of local U(N)$_{\rm L}\otimes$ global U(N)$_{\rm R}$ fields.  In section 3 we
discuss the BRS symmetry, which becomes very transparent in the present
formulation, for a SU(N) theory.  In the remaining sections the BRS symmetry
and the pinching technique of Cornwall \cite{corn} are used to perform the
perturbative renormalization, which becomes much more tractable in the present
formulation, of the effective theory.  This culminates in our main result,
namely, that in the perturbative domain the present quantization procedure
corresponds to Landau gauge fixing for non-abelian theories.  The normal
perturbative results are therefore recovered, showing that the Gribov problem
has no effect in the perturbative domain.  Some technical results are collected
in two appendices.

\section{The U(N) formalism}

In \cite{st} the auxiliary bosonic and fermionic fields were taken as vector
valued in the fundamental representation of SU(N).  To integrate the
gauge degrees of freedom out in this setting one has to resort to a sequential
procedure in which the gauge symmetry is broken down according to
SU(N)$\supset$SU(N-1)$\ldots$.  This procedure obscures many aspects of the
theory and leads to technical complications.

Here we avoid this sequential procedure by taking the auxiliary bosonic and
fermionic fields to be matrix valued in the fundamental representation of U(N).
We exploit, as a matter of convenience, the fact that SU(N)$\subset$U(N).
Our procedure is as follows: consider a pure Yang--Mills type theory (for
notation see appendix A)

\begin{eqnarray}\label{2.1}
L_{\rm YM} &=& - \frac{1}{2}{\rm tr}(G^2_{\mu \nu}) \nonumber\\[-2mm]
 & & \\[-2mm]
G_{\mu \nu} &=&\partial_{\mu} G_{\nu} - \partial_{\nu} G_{\mu} + ig [G_{\mu},
G_{\nu}] \; , \quad G_{\mu} = G^a_{\mu} t_a \nonumber \end{eqnarray}

This lagrangian is invariant under the gauge transformations

\begin{equation}\label{2.2}
G_{\mu} \rightarrow G'_{\mu} = - \frac{i}{g} U D_{\mu} U^\dagger \; , \quad
U\in {\rm U(N)} \end{equation}
where

\begin{equation}
D_{\mu} = \partial_{\mu} + ig G_{\mu}
\end{equation}
is the gauge covariant derivative in the fundamental representation.  Now for
some set of gauge invariant functionals $O[G] = O[G']$ define

\begin{eqnarray}\label{2.4}
<O> &=& Z^{-1} \int [dG]O[G] \, \mbox {\rm exp} ~(i \int d^4 x L_{\rm YM})
\; , \nonumber\\[-2mm]
 & & \\[-2mm]
Z &=& \int [dG] \, \mbox {\rm exp} ~(i \int d^4 x L_{\rm YM}) \nonumber
\end{eqnarray}

For the restricted class of functionals, \underline{$O$}, which depend on
the SU(N) gauge field

\begin{equation}
\underline{G}_{\mu} = G^{\underline a}_{\mu} t_{\underline{a}}
\end{equation}
only, it follows from U(N)$\simeq$ SU(N)$\otimes$U(1) that
\begin{equation}
\begin{array}{rcl} < \underline{O} > &=& \underline{Z}^{-1} \displaystyle
\int [d\underline{G}]\, \underline{O} [\underline{G}] \, \mbox {\rm exp} ~(i
\displaystyle \int d^4 x L_{\rm YM}) \; , \\[5mm] \underline{Z} &=&
\displaystyle
\int [d\underline{G}] \, \mbox {\rm exp} ~(i \displaystyle \int d^4 x L_{\rm
YM})
\end{array} \label{2.4'}
\end{equation}
which is canonical.  Thus quantization of an SU(N) gauge theory proceeds
from (\ref{2.4}) applied to $\underline{O}$.

While (\ref{2.4}) and (\ref{2.4'}) are well defined on the lattice, in the
continuum the measure
\begin{equation}
[dG] \propto \Pi_{x,\mu,a} d G^a_{\mu} (x)
\end{equation}
is itself gauge invariant, $[dG'] = [dG]$, so $< \underline{O}> \sim
\infty/\infty$ which is ill defined.  To factor the volume of the gauge group
while eschewing the Faddeev--Popov ansatz together with its associated Gribov
problem consider the (minimal) identity

\begin{eqnarray}\label{2.6} 1 &=& \int [d \Phi] [d \Phi^\dagger] [d \zeta] [d
\zeta^\dagger] \, \mbox {\rm exp} ~(i \int d^4 x L_{{\rm aux}}) \; ,
\nonumber\\[-2mm] & & \\[-2mm] L_{{\rm aux}} &=& {\rm tr} (\Phi^\dagger
D_{\mu}^\dagger D^{\mu} \Phi + \zeta^\dagger D^\dagger_{\mu} D^{\mu} \zeta)\,.
\nonumber \end{eqnarray}
Herein the auxiliary fields are matrix valued in the fundamental representation
of U(N)

\begin{equation}
\Phi = \Phi_a t_a \; , \quad \Phi^\dagger = \Phi^\dagger_a t_a\quad{\rm
and}\quad \zeta = \zeta_a t_a \; , \quad \zeta^\dagger = \zeta^\dagger_a t_a
\end{equation}
representing $N^2$ complex ($2N^2$ real) scalar degrees of freedom and
$N^2$ complex Grassmann valued scalar degrees of freedom (ghosts),
respectively.  Furthermore $D^\dagger_{\mu}$ acts to the left

\begin{equation}
D^\dagger_{\mu} = \stackrel{\leftarrow}{\partial}_{\mu} - i g G_{\mu} \, .
\end{equation}

The measure on the auxiliary fields is

\begin{equation}
[d \Phi] [d \Phi^\dagger] [d \zeta] [d \zeta^\dagger] = \Pi_{x, a}
\frac{d \Phi_a
(x) d \Phi^\dagger_a (x)}{2 \pi i} d \zeta_a (x) d \zeta^\dagger_a (x) \, .
\end{equation}

By construction $L_{{\rm aux}}$ has local U(N)$_{\rm L}$ invariance, under the
transformations

\begin{equation}\label{2.11}
\begin{array}{rclclcl}
G_{\mu} &\longrightarrow& - \frac{i}{g} U_{\rm L} D_{\mu} U^\dagger_{\rm L} \;
,
\quad \Phi &\longrightarrow& U_{\rm L} \Phi \; , \quad \Phi^\dagger
&\longrightarrow&
\Phi^\dagger U^\dagger_{\rm L} \; ,\\[2mm]
\zeta &\longrightarrow& U_{\rm L}\zeta\;,\qquad\qquad \zeta^\dagger
&\longrightarrow& \zeta^\dagger
U^\dagger_{\rm L} \; , \quad U_{\rm L} &=& \mbox {\rm exp} ~(i \theta^{\rm L}_a
t_a) \in {\rm U(N)}_{\rm L} \end{array}
\end{equation}
as well as an independent global U(N)$_{\rm R}$ symmetry:

\begin{equation}\label{2.12}
\begin{array}{rclclcl}
G_{\mu} &\longrightarrow& G_{\mu} \; , \quad \Phi &\longrightarrow& \Phi
U^\dagger_{\rm R} \; , \quad \Phi^\dagger &=& U_{\rm R} \Phi^\dagger \; ,
\\[2mm]
\zeta &\longrightarrow& \zeta U^\dagger_{\rm R} \; , \quad \zeta^\dagger
&\longrightarrow& U_{\rm R}
\zeta^\dagger \; , \quad U_{\rm R} &=& \mbox {\rm exp} ~(i \theta^{\rm R}_a
t_a) \in {\rm U(N)}_{\rm R} \, . \end{array}
\end{equation}

For infinitesimal transformations

\begin{equation}
\delta_{\rm L} \Phi_a = 2 i {\rm tr} (t_a t_c t_b) \theta^{\rm L}_c \Phi_b = i
\theta^{\rm L}_c (T_c)_{ab} \Phi_b \end{equation}
whereas

\begin{equation}
\delta_{\rm R} \Phi_a = i \theta^{\rm R}_c (- T^{\ast}_c)_{ab} \Phi_b
\end{equation}
which, together with (\ref{A.14}), identifies $T_a$ and $-
T_a^{\ast}$
as generators of the adjoint representation of U(N)$_{\rm L}$ and U(N)$_{\rm
R}$ respectively; our auxiliary fields transform as the basis for the adjoint
representation of chiral U(N).  Note that this is in contrast to
\cite{st} where the auxiliary fields transform as the fundamental
representation.  Writing

\begin{eqnarray}
\int d^4 x L_{{\rm aux}} &=& \int d^4 x \int d^4 y \left(\frac{1}{2}
\Phi^\dagger_a
(x) M_{ab} (x, y) \Phi_b (y) + \frac{1}{2} \zeta^\dagger_a (x) M_{ab} (x, y)
\zeta (y)\right) \; , \nonumber\\[-2mm]
 & & \\[-2mm]
&&M_{ab} (x, y) = - D^{ae}_{\mu} D^{\mu}_{eb} \delta^{(4)} (x - y) \nonumber
\end{eqnarray}
with

\begin{equation}
D^{ab}_{\mu} = \delta_{ab} \partial_{\mu} + ig G^e_{\mu} (T_e)_{ab}
\end{equation}
the gauge covariant derivative in the adjoint representation of U(N)$_{\rm L}$,
the proof of the identity is immediate:
\begin{equation}
\int [d \Phi] [d \Phi^\dagger] [d \zeta] [d \zeta^\dagger] \, \mbox {\rm exp}
{}~(i
\int d^4 x L_{{\rm aux}}) = \frac{{\rm det} ~(M)}{{\rm det} ~(M)} = 1 \, .
\end{equation}
Injecting (\ref{2.6}) into (\ref{2.4})

\begin{eqnarray}
< \underline{O} > &=& Z^{-1} \int [d \Phi] [d \Phi^\dagger] [d \zeta] [d
\zeta^\dagger]
[dG] \underline{O} [\underline{G}] \, \mbox {\rm exp} ~(i \int d^4 x (L_{\rm
YM} +
L_{{\rm aux}})) \; , \nonumber\\[-2mm]
 & & \\[-2mm]
Z &=& \int [d \Phi] [d \Phi^\dagger] [d \zeta] [d \zeta^\dagger] [dG] \, \mbox
{\rm exp} ~(i\int d^4 x (L_{\rm YM} + L_{{\rm aux}})) \nonumber \end{eqnarray}
we make the change of variables

\begin{eqnarray}\label{2.18}
\Phi \; = \; - U (\pi) \phi &,& \Phi^\dagger \; = \; - \phi U^\dagger (\pi)
\nonumber\\[-2mm]
 & & \\[-2mm]
U (\pi) &=& \mbox {\rm exp} ~(i \pi_a t_a) \nonumber
\end{eqnarray}
followed by

\begin{equation}\label{2.19}
G_{\mu} \longrightarrow G'_{\mu} = - \frac{i}{g} U (\pi) D_{\mu} U^\dagger
(\pi) \;
, \quad \zeta \longrightarrow \zeta' = U (\pi) \zeta \; , \quad \zeta^\dagger
\longrightarrow \zeta'^\dagger = \zeta^\dagger U^\dagger (\pi) \end{equation}
which is the form of a (unitary) gauge transformation.  Using the result
(\ref{B.9})

\begin{equation} \begin{array}{rcl} < \underline{O} > &=& Z^{-1} \displaystyle
\int [dU] [{\rm det} ~(J (\phi)) d \phi] [d \zeta] [d \zeta^\dagger] [dG]\,
\underline{O} [\underline{G}] \, \mbox {\rm exp} ~(i \displaystyle \int d^4 x
L_{\rm eff})\\[5mm] Z &=& \displaystyle \int [dU] [{\rm det} ~(J (\phi)) d
\phi] [d \zeta] [d \zeta^\dagger] [dG] \, \mbox {\rm exp} ~(i \displaystyle
\int d^4 x L_{\rm eff})\label{2.19a} \end{array} \end{equation}
where

\begin{equation}
L_{\rm eff} = L_{\rm YM} + {\rm tr} (\phi D_{\mu}^\dagger D^{\mu} \phi +
\zeta^\dagger
D^\dagger_{\mu} D^{\mu} \zeta) \end{equation}
and

\begin{equation}\label{2.21}
J_{ab} (\phi) = D_{abc} \phi_c
\end{equation}

The integrand being independent of $U (\pi)$ the volume of the gauge group,
$\displaystyle \int [dU]$, factors and cancels in the normalization, leaving

\begin{eqnarray} < \underline{O} > &=& Z^{-1} \int [{\rm det} ~(J (\phi)) d
\phi] [d \zeta] [d \zeta^\dagger] [dG]\, \underline{O} [\underline{G}] \, \mbox
{\rm exp} ~(i \int d^4 x L_{\rm eff}) \nonumber\\[-2mm] & & \\[-2mm] Z &=& \int
[{\rm det} ~(J (\phi)) d \phi] [d \zeta] [d \zeta^\dagger] [dG] \, \mbox {\rm
exp} ~(i \int d^4 x L_{\rm eff}) \nonumber \end{eqnarray}
Comparing our formulation to that of \cite{zwan,jl} one sees that the latter
constitutes a non-linear realization of the chiral symmetry with the ghosts
represented by pseudofermions.

\section{BRS symmetry}

The essential content of the identity (\ref{2.6}) is that $G_{\mu}$ appears as
an external source field, expanding in powers of $G^a_{\mu}$ one observes that
for each closed $\Phi-$loop there is a closed $\zeta-$loop with a relative
minus sign from ghost statistics so they exactly cancel -- including vacuum
graphs -- much as in a supersymmetry.  In turn this suggests that our procedure
possesses a type of BRS symmetry, and such is indeed the case as we now proceed
to show.  For the representation of the auxiliary fields used here the relevant
(anti) BRS transformations generated by $(\bar{S}) S$ are

\begin{equation}\label{3.1}
\begin{array}{rclcrcl}
S \Phi^\dagger &=& \zeta^\dagger & \hspace*{3cm} & \bar{S} \Phi^\dagger &=&
0\\[3mm] S \Phi &=& 0 & & \bar{S} \Phi &=& - \zeta\\[3mm]
S \zeta^\dagger &=& 0 & & \bar{S} \zeta^\dagger &=& \Phi^\dagger\\[3mm]
S \zeta &=& \Phi & & \bar{S} \zeta &=& 0\\[3mm]
S G_{\mu} &=& 0 & & \bar{S} G_{\mu} &=& 0
\end{array}
\end{equation}
Clearly $S$ and $\bar{S}$ are nilpotent but do not anticommute, rather

\begin{equation}
(S \bar{S} + \bar{S} S) (\Phi^\dagger \; , \; \Phi \; , \; \zeta^\dagger \ , \;
\zeta) = (\Phi^\dagger \; , \; - \Phi \; , \; \zeta^\dagger \; , \; - \zeta)
\end{equation}
Using
\begin{equation}
S (XY) = (SX) Y \pm X (SY) \; , \quad \bar{S} (XY) = (\bar{S}X) Y \pm X
(\bar{S}Y)
\end{equation}
where the + ($-$) sign applies if $X$ is ghost even (odd) one sees that

\begin{equation}\label{3.3}
L_{{\rm aux}} = S W = \bar{S} W^{\ast}
\end{equation}
with

\begin{equation}
W = {\rm tr} (\Phi^\dagger D^\dagger_{\mu} D^{\mu} \zeta) \; , \quad W^{\ast} =
{\rm tr}(\zeta^\dagger D^\dagger_{\mu} D^{\mu} \Phi) \, . \end{equation}

Thus from nilpotency $L_{{\rm aux}}$ is (anti) BRS invariant and moreover a
simple calculation shows the BRS invariance of the measure in (\ref{2.6}).
Then, for $F$ any functional of the auxiliary fields and $G_{\mu}$
\begin{eqnarray}
&&\int [d \Phi] [d \Phi^\dagger] [d \zeta] [d \zeta^\dagger]\,F
\mbox {\rm exp}
{}~(i \int d^4 x L_{{\rm aux}}) =\nonumber\\&&\int [d \Phi] [d \Phi^\dagger] [d
\zeta] [d
\zeta^\dagger] (F + \chi SF) \, \mbox {\rm exp} ~(i \int d^4 x L_{{\rm aux}})
\end{eqnarray}
where we have performed a BRS transformation with $\chi$ a global
Grassmann variable.  It follows that

\begin{equation}\label{3.5}
0 = \int [d \Phi] [d \Phi^\dagger] [d \zeta] [d \zeta^\dagger] (SF) \, \mbox
{\rm exp} ~(i \int d^4 x L_{{\rm aux}}) \end{equation}

{}From (\ref{3.5}) we have an alternative proof of (\ref{2.6}):  let

\begin{equation}
I [G] = \int [d \Phi] [d \Phi^\dagger] [d \zeta] [d \zeta^\dagger] \, \mbox
{\rm exp} ~(i \int d^4 x L_{{\rm aux}}) \, . \end{equation}

By (\ref{3.3}), differentiation of $I[G]$ with respect to $G_{\mu}$ produces a
quantity of the form (\ref{3.5}), i.e. $\delta I [G] / \delta G_{\mu}^a = 0$
which is
to say $I [G]$ is a constant that may be normalized to unity.  Note from
(\ref{3.3}) that $L_{\rm aux}$ corresponds to a topological field theory
of the Witten type \cite{bir}.

Next we need to establish what becomes of this BRS symmetry under the change of
variables (\ref{2.18}), (\ref{2.19}).  We begin by defining the `BRS current'.

\begin{eqnarray}\label{3.7}
C &=& U^\dagger (\pi) S U (\pi) \nonumber\\[-2mm]
 & & \\[-2mm]
SU(\pi) \; = \; U (\pi) C &,& SU^\dagger(\pi) \; = \; - CU^\dagger (\pi)
\nonumber \end{eqnarray}
Then

\begin{equation}S (- \phi U^\dagger (\pi)) = - (S \phi) U^\dagger (\pi) + \phi
C U^\dagger (\pi) = \zeta^\dagger U^\dagger (\pi)\end{equation}
or

\begin{equation}\label{3.8}
S \phi = - \zeta^\dagger + \phi C
\end{equation}
while

\begin{equation}S (U (\pi) \phi) = U (\pi) C \phi + U (\pi) (S
\phi)\end{equation}
so

\begin{equation}\label{3.9}
S \phi = - C \phi\,.
\end{equation}

Similarly

\begin{eqnarray*}
S (\zeta^\dagger U^\dagger (\pi)) &=& (S \zeta^\dagger) U^\dagger (\pi) +
\zeta^\dagger C U^\dagger (\pi)) \; = \; 0\\
S (U (\pi) \zeta) &=& U (\pi) C \zeta + U (\pi) (S \zeta)) \; = \; - U (\pi)
\phi
\end{eqnarray*}
yielding

\begin{eqnarray}
S \zeta^\dagger &=& - \zeta^\dagger C\\
S \zeta &=& - \phi - C \zeta
\end{eqnarray}

Finally (note $S G'_{\mu} = 0$ now)

\begin{eqnarray*} S G_{\mu} &=& S \left(- \frac{i}{g} U^\dagger (\pi)
(\partial_{\mu}
+ i g G'_{\mu}) U (\pi)\right)\\ &=& - \frac{i}{g} (- C U^\dagger (\pi))
(\partial_{\mu} + i g G'_{\mu}) U (\pi) - \frac{i}{g} U^\dagger (\pi)
(\partial_{\mu} + i g G'_{\mu}) (U (\pi) C)\\ &=& - C G_{\mu} + G_{\mu} C -
\frac{i}{g} \partial_{\mu} C \end{eqnarray*}
so

\begin{equation}
S G_{\mu} = -\frac{i}{g} [D_{\mu}, C]
\end{equation}

What is most crucial is that consistency between (\ref{3.8}) and (\ref{3.9})
requires

\begin{equation}
\zeta^\dagger = \{\phi, C\}
\end{equation}
or in component form

\begin{equation}
\zeta^\dagger_a = D_{abe} \phi_e C_b = J_{ab} (\phi) C_b
\end{equation}
where $J$ is the same matrix whose determinant appears in the measure in
(\ref{2.19a}).  Thus, changing variables from $\zeta^\dagger$ to $C$

\begin{eqnarray}\label{3.15} < \underline{O} > &=& Z^{-1} \int [dG] [d \phi] [d
\zeta]
[dC] \underline{O} [\underline{G}] \, \mbox {\rm exp} ~(i \int d^4 x L_{\rm
eff})
\nonumber\\ Z &=& \int [dG] [d \phi] [d \zeta] [dC] \, \mbox {\rm exp} ~(i \int
d^4 x L_{\rm eff})\\ L_{\rm eff} &=& L_{\rm YM} + {\rm tr} (\phi
D^\dagger_{\mu} D_{\mu}
\phi + \{\phi, C\} D^\dagger_{\mu} D^{\mu} \zeta) \nonumber \end{eqnarray}

The BRS transformations are now

\begin{eqnarray}\label{3.16}
S \phi &=& - C \phi \nonumber\\
SC &=& - CC \nonumber\\[-2mm]
 & & \\[-2mm]
S \zeta &=& - \phi - C \zeta \nonumber\\
S G_{\mu} &=& - \frac{i}{g} [D_{\mu}, C] \nonumber
\end{eqnarray}
where that for $C$ follows directly from the definition (\ref{3.7}) and
nilpotency of $S$ -- it is a straightforward exercise to show the nilpotency of
the BRS transformations (\ref{3.16}) and that

\begin{equation}
L_{\rm eff} = L_{\rm YM} + S {\rm tr} (- \phi D^\dagger_{\mu} D^{\mu} \zeta) \,
{}.
\end{equation}

As $SG_{\mu}$ has the form of a gauge transformation BRS invariance of
$L_{\rm eff}$ is immediate.  Moreover, again a simple calculation demonstrates
that
the measure in (\ref{3.15}) is BRS invariant; notable in this regard is that
the $G_{\mu}$ and $C$, $\phi$ and $\zeta$ contributions to the
superdeterminant cancel pairwise.  In turn this is understood in that the
change of variables (\ref{2.18}), (\ref{2.19}) absorbs the auxiliary field
$\pi$ into the gauge field.

A parallel development may be made for the fate of the anti BRS
symmetry. With $U^\dagger(\pi)\bar S U(\pi)\equiv C^\dagger$:

\begin{eqnarray}
\bar S \phi &=& \phi C^\dagger=\zeta-C^\dagger\phi \nonumber\\
\bar S C^\dagger &=& - C^\dagger C^\dagger \nonumber\\[-2mm]
 & & \\[-2mm]
\bar S \zeta^\dagger &=& - \phi - \zeta^\dagger C^\dagger\\
\bar S \zeta &=& - C^\dagger\zeta \nonumber\\
\bar S G_{\mu} &=& - \frac{i}{g} [D_{\mu}, C^\dagger] \nonumber
\end{eqnarray}

Substituting $(\zeta^\dagger)$ $\zeta$ in terms of $\phi$ and $(C)$ $C^\dagger$
in the corresponding (anti) BRS transformations one obtains

\begin{equation}
\{\phi,SC^\dagger\}+\{C,C^\dagger\}\phi=-\phi=\{\phi,\bar
SC\}+\phi\{C,C^\dagger\}
\end{equation}
so

\begin{equation}
\{\phi,SC^\dagger+\bar S C+CC^\dagger+C^\dagger C\}=-2\phi
\end{equation}
with solution

\begin{equation}\label{3.21}
SC^\dagger+\bar S C+\{C,C^\dagger\}=-I
\end{equation}
or in component form

\begin{equation}
SC^\dagger_a+\bar S C_a+if_{abc}C_bC^\dagger_c=-\sqrt{2N}\delta_{a0}\,.
\end{equation}

It is then easy to verify that the BRS and anti BRS transformations now do
satisfy both nilpotency and anti commutation

\begin{equation}
S^2=S\bar S+\bar S S=\bar S^2=0
\end{equation}
as well as that

\begin{equation}
L_{\rm eff}=L_{\rm YM}+\bar SS{\rm tr}(\phi D^\dagger_\mu D^\mu\phi)\,.
\end{equation}
The essential difference between our BRS algebra and the usual one \cite{brs}
lies in the nonvanishing right hand side of (\ref{3.21})

Transformation from the $\zeta$ to $C^\dagger$ ghost in (\ref{3.15}) leads to
the occurrence of $1/\det(J(\phi))$ in the measure while exacerbating the
problem of higher than quartic vertices in $L_{\rm eff}$, however, and thus we
implement only the BRS symmetry in what follows.

\section{Extensions}

The identity (\ref{2.6}) is not the most general such we could write down; any
$L_{{\rm aux}}$ possessing the chiral symmetry (\ref{2.11}), (\ref{2.12}) and
the BRS symmetry (\ref{3.1}) will do.  If on the other hand, we impose the
condition of perturbative renormalizability (in the sense of orthodox gauge
fixing) and require that $L_{{\rm aux}}$ retain the scale invariance of
$L_{\rm YM}$ then the most general form is

\begin{eqnarray}\label{4.1} 1 &=& \int [d \Phi] [d \Phi^\dagger] [d \zeta] [d
\zeta^\dagger] \, \mbox {\rm exp} ~(i \int d^4 x L_{{\rm aux}}) \; ,
\nonumber\\[-2mm] & & \\[-2mm] L_{{\rm aux}} &=& {\rm tr} \left(\Phi^\dagger
D_{\mu}^\dagger
D^{\mu} \Phi + \zeta^\dagger D^\dagger_{\mu} D^{\mu} \zeta -
\frac{\lambda_1}{N} (\Phi^\dagger
\Phi + \zeta^\dagger \zeta)^2 \right) - \frac{\lambda_2}{N^2} ({\rm tr}
(\Phi^\dagger \Phi + \zeta^\dagger \zeta))^2 \nonumber \end{eqnarray}

This may be proven by uncompleting the square:  let

\begin{eqnarray} I [G] &=& {\cal N}^{-1} \int [d \Phi] [d \Phi^\dagger] [d
\zeta] [d \zeta^\dagger] [d \Sigma] \, \mbox {\rm exp} ~(i \int d^4 x
(\tilde{L}_{{\rm aux}} + L_{\Sigma}) \; , \nonumber\\ {\cal N} &=& \int [d
\Sigma]
\, \mbox {\rm exp} ~(i \int d^4 x L_{\Sigma}) \; , \nonumber\\[-2mm] & &
\\[-2mm]
\tilde{L}_{{\rm aux}} &=& {\rm tr} (\Phi^\dagger D_{\mu}^\dagger D^{\mu} \Phi +
\zeta^\dagger
D_{\mu}^\dagger D^{\mu} \zeta - (\Phi^\dagger \Phi + \zeta^\dagger \zeta)
\Sigma) \; , \; \Sigma =
\Sigma_a t_a \; , \nonumber\\ L_{\Sigma} &=& \frac{N}{4 \lambda_1}{\rm
tr}(\Sigma^2)
- \frac{1}{4 \lambda_1} \frac{\lambda_2}{\lambda_1 + \lambda_2} ({\rm tr}
\Sigma)^2 \nonumber \end{eqnarray}
Integrating out the $\Sigma-$field gives

\begin{equation} I [G] = \int [d \Phi] [d \Phi^\dagger] [d \zeta] [d
\zeta^\dagger] \, \mbox {\rm exp} ~(i \int d^4 x L_{{\rm aux}}) \end{equation}
with $L_{{\rm aux}}$ as in (\ref{4.1}).  Conversely, integrating out the
auxiliary fields $\Phi$, $\Phi^\dagger$, $\zeta$ and $\zeta^\dagger$ first,
with now

\begin{equation} M_{ab} (x, y) = - (D^{ae}_{\mu} D^{\mu}_{eb} + \Sigma_e
(T^{\ast}_e)_{ab}) \delta^{(4)} (x - y) \end{equation}
yields

\begin{equation} I [G] = {\cal N}^{-1} \int [d \Sigma] \frac{{\rm det}
{}~(M)}{{\rm det} ~(M)} \, \mbox {\rm exp} ~(i \int d^4 x L_{\Sigma}) = 1
\end{equation}

Alternately one may observe that

\begin{eqnarray}
L_{{\rm aux}} &=& SW \; , \nonumber\\[-2mm]
 & & \\[-2mm]
W &=&{\rm tr}\left(\Phi^\dagger D^\dagger_{\mu} D^{\mu} \zeta -
\frac{\lambda_1}{N} \Phi^\dagger \zeta
S (\Phi^\dagger \zeta)\right) - \frac{\lambda_2}{N^2}{\rm tr}(\Phi^\dagger
\zeta) S{\rm tr} (\Phi^\dagger \zeta) \nonumber \end{eqnarray}
so (\ref{3.5}) holds here also: differentiation of $I [G]$ by $G_{\mu}^a$,
$\lambda_1$ and $\lambda_2$ yields a vanishing quantity, $I [G]$ thereby being
a constant which may be normalized to unity.  Further, clearly, one may again
follow the steps leading to (\ref{3.15}), but now with

\begin{eqnarray}\label{4.7} L_{\rm eff} &=& L_{\rm YM} + SW \; ,
\nonumber\\[-2mm] & &
\\[-2mm]
W &=& {\rm tr} \left(- \phi D^\dagger_{\mu} D^{\mu} \zeta - \frac{\lambda_1}{N}
\phi \zeta S (\phi \zeta)\right) - \frac{\lambda_2}{N^2}{\rm tr}(\phi \zeta) S
{\rm tr} (\phi \zeta) \nonumber \end{eqnarray}

Due to the BRS invariance of $L_{\rm eff}$ and the measure in (\ref{3.15}) we
have
analogous to (\ref{3.5})

\begin{equation}\label{4.8} 0 = \int [dG] [d \phi] [d \zeta] [dC] (SF) \, \mbox
{\rm exp} ~(i \int d^4 x L_{\rm eff}) \end{equation}
for $F$ any functional of $G_{\mu}$ and the auxiliary fields $\phi$, $\zeta$
and $C$.  From (\ref{4.7}) and (\ref{4.8})

\begin{equation} \frac{\partial}{\partial \lambda_1} Z =
\frac{\partial}{\partial \lambda_2} Z = \frac{\partial}{\partial \lambda_1} <
\underline{O} > = \frac{\partial}{\partial \lambda_2} <
\underline{O} > = 0 \end{equation}

One should take care to note that the independence of the `partition function'
$Z$ and $< O >$ from $\lambda_1$ and $\lambda_2$ does not mean that we
should blindly set $\lambda_1 = \lambda_2 = 0$.  Indeed, a cursory examination
of the system $L_{\rm YM} + L_{{\rm aux}}$ treated with perturbative gauge
fixing
demonstrates the generation of chiral invariant quartic interactions among the
auxiliary fields which require the additional terms in (\ref{4.1}) for their
renormalization.

Thus far we have restricted ourselves to pure Yang--Mill theory, but for
applications to, in particular QCD, and also for the purpose of discussions to
follow we need the extension of our formalism to SU(N) gauge fields coupled
to matter.  Consider therefore

\begin{equation}
L = L_{\rm YM} + \bar{\psi} (i \gamma^\mu\underline{D}_\mu - m) \psi
\end{equation}
where $L_{\rm YM}$ is as in (\ref{2.1}), and

\begin{equation}
\underline{D}_{\mu} = \partial_{\mu} + ig \underline{G}_{\mu}
\end{equation}
with $m$ the (bare) fermion mass or, for $N_f$ flavors, the (diagonal) mass
matrix.  This lagrangian is invariant under (\ref{2.2}) together with

\begin{equation} \psi \longrightarrow \underline{U} \psi \; , \quad \bar{\psi}
\longrightarrow \bar{\psi} \underline{U}^\dagger \; , \quad \underline{U}
\in {\rm U(N)/U(1)} \end{equation}

The gauge invariant observables to be considered are $\underline{O} =
\underline{O} [\underline{G}, \, \bar{\psi}, \, \psi]$ ,

\begin{eqnarray} < \underline{O} > &=& Z^{-1} \int [d \psi] [d \bar{\psi}] [dG]
\underline{O} [\underline{G}, \, \bar{\psi}, \, \psi] \, \mbox {\rm exp} ~(i
\int d^4 x L) \; , \nonumber\\[-2mm] & & \\[-2mm] Z &=& \int [d \psi] [d
\bar{\psi}] [dG] \, \mbox {\rm exp} ~(i \int d^4 x L) \nonumber \end{eqnarray}

Supplementing the BRS transformations (\ref{3.1}) by

\begin{equation}
S \psi = S \bar{\psi} = 0
\end{equation}
and the change of variables (\ref{2.19}) with

\begin{eqnarray} \psi \longrightarrow \psi' &=& \underline{U} (\pi) \psi \; ,
\quad \bar{\psi} \longrightarrow \bar{\psi}' \; = \; \bar{\psi}
\underline{U}^\dagger
(\pi) \; , \nonumber\\[-2mm] & & \\[-2mm] \underline{U} (\pi) &=& \mbox {\rm
exp} ~(i \pi_{\underline{a}} t_{\underline{a}}) \nonumber \end{eqnarray}
one straightforwardly arrives at

\begin{eqnarray}\label{4.16} < \underline{O} > &=& Z^{-1} \int [d \psi] [d
\bar{\psi}] [dG] [d \phi] [d \zeta] [dC]\,\underline{O} [\underline{G}, \,
\bar{\psi}, \, \psi] \, \mbox {\rm exp} ~(i \int d^4 x L_{\rm eff}) \; ,
\nonumber\\[-2mm] & & \\[-2mm] Z &=& \int [d \psi] [d \bar{\psi}] [dG] [d \phi]
[d \zeta] [dC] \, \mbox {\rm exp} ~(i \int d^4 x L_{\rm eff}) \nonumber
\end{eqnarray} where

\begin{equation}
L_{\rm eff} = L + SW
\end{equation}
and

\begin{equation} W = {\rm tr} \left(- \phi D^\dagger_{\mu} D^{\mu} \zeta -
\frac{\lambda_1}{N} \phi \zeta S (\phi \zeta)\right) - \frac{\lambda_2}{N^2}
{\rm tr} (\phi \zeta) S {\rm tr} (\phi \zeta) \end{equation}
while the BRS transformations are

\begin{eqnarray}\label{4.19} S \phi &=& - C \phi \nonumber\\ SC &=& - CC
\nonumber\\
S \zeta &=& - \phi - C \zeta \nonumber\\[-3mm] & & \\[-3mm] SG_{\mu} &=& -
\frac{i}{g} [D_{\mu}, C] \nonumber\\ S \psi &=& - \underline{C} \psi \; , \quad
\underline{C} \; = \; C_{\underline{a}} t_{\underline{a}} \nonumber\\ S
\bar{\psi} &=& \bar{\psi} \underline{C}\,. \nonumber \end{eqnarray}
(note the matter fields are ghost even).  By BRS invariance of $L_{\rm eff}$
and
the measure in (\ref{4.16})

\begin{equation} 0 = \int [d \psi] [d \bar{\psi}] [dG] [d \phi] [d \zeta] [dC]
(SF) \, \mbox {\rm exp} ~(i \int d^4 x L_{\rm eff}) \end{equation}
for any functional $F$ of $\psi$, $\bar{\psi}$, $G_{\mu}$ and the auxiliary
fields $\phi$, $\zeta$ and $C$.

\section{Renormalization}

The intrinsic BRS symmetry of our quantization prescription imposes some
important constraints on the renormalizations which will be needed below to
deal with infinities.  In particular, preservation of (\ref{3.1}) for the
renormalized fields requires a common wavefunction renormalization constant:

\begin{equation}
(\Phi^\dagger_B \; , \; \Phi_B \; , \; \zeta^\dagger_B \; , \; \zeta_B) =
(\tilde{Z})^{1/2} (\Phi^\dagger_R \; , \; \Phi_R \; , \; \zeta^\dagger_R \; ,
\; \zeta_R) \end{equation}
wherein $B (R)$ denotes bare (renormalized).  Also, in view of the remarks in
section 4, we write for the renormalization of the quartic couplings

\begin{equation}
\lambda_{_Bi} = \delta \lambda_i / (\tilde{Z})^2 \; ,
\end{equation}
i.e., we hold the renormalized coupling constants $\lambda_{_Ri}$ to vanish.

For the gauge field and its coupling we take

\begin{eqnarray}
G_{_B \mu} &=& (Z_3)^{1/2} G_{_R\mu} \nonumber\\[-2mm]
 & & \\[-2mm]
g_{_B} &=& Z_g g_{_R} \nonumber
\end{eqnarray}

Carrying out the transformations (\ref{2.18}), (\ref{2.19}) with $U_B (\pi) =
U_{\rm R} (\pi) = U (\pi)$ we have

\begin{eqnarray}
\phi_B &=& (\tilde{Z})^{1/2} \phi_R \nonumber\\[-2mm]
 & & \\[-2mm]
C_B &=& C_R \; = \; C \nonumber
\end{eqnarray}

Note the non-renormalization of $C$, as follows from its definition (\ref{3.7})
as a symmetry current.  Defining the renormalized covariant derivative

\begin{equation}
D_{_R\mu} = \partial_{\mu} + i Z_g \sqrt{Z_3}\, g_{_R} G_{_R\mu}
\end{equation}
one immediately obtains the renormalized BRS transformations

\begin{eqnarray}\label{5.6}
S \phi_R &=& - C \phi_R \nonumber\\
SC &=& - CC \nonumber\\[-2mm]
 & & \\[-2mm]
S \zeta_R &=& - \phi_R - C \zeta_R \nonumber\\
SG_{_R\mu} &=& - \frac{i}{Z_g \sqrt{Z_3}} g_{_R} [D_{_R\mu}, C] \nonumber
\end{eqnarray}

That (\ref{5.6}) are indeed the BRS symmetry transformations of

\begin{eqnarray} L^B_{\rm eff} &=& - \frac{Z_3}{2}{\rm tr}(G^2_{_R\mu\nu}) +
\tilde{Z}{\rm tr}(\phi_R D^\dagger_{_R \mu} D^{\mu}_{_R} \phi_R + \{\phi_R, C\}
D^\dagger_{_R\mu} D^{\mu}_{_R} \zeta_R ) -\nonumber\\ &&\frac{\delta
\lambda_1}{N}{\rm tr} ((\phi^2_R + \{\phi_R, C\} \zeta_R)^2) - \frac{\delta
\lambda_2}{N^2} ({\rm tr} (\phi^2_R + \{\phi_R, C\} \zeta_R))^2 \; , \\ G_{_R
\mu \nu} &=& \partial_{\mu} G_{_R \nu} - \partial_{\nu} G_{_R \mu} + i Z_g
\sqrt{Z_3}\, g_{_R} [G_{_R \mu}, G_{_R \nu}] \nonumber \end{eqnarray}
for the pure gauge theory is by inspection.

Including fermions, we need the additional field and mass renormalizations

\begin{eqnarray}
(\bar{\psi}_B, \, \psi_B) &=& (Z_2)^{1/2} (\bar{\psi}_R, \, \psi_R)
\nonumber\\[-2mm]
 & & \\[-2mm]
m_{_B} &=& m_{_R} Z_m/Z_2 \nonumber
\end{eqnarray}
and then

\begin{eqnarray} L^B_{\rm eff} &=& \bar{\psi}_R (Z_2 i
\gamma^\mu\underline{D}_{_R\mu}
- m_{_R} Z_m) \psi_R - \frac{Z_3}{2}{\rm tr}(G^2_{_R \mu \nu}) + SW_B \; ,
\nonumber\\[-2mm] & & \\[-2mm] W_B &=& \tilde{Z}{\rm tr}(- \phi_R D^\dagger_{_R
\mu}D^\mu_{_R}
\zeta_R) - \frac{\delta \lambda_1}{N}{\rm tr}(\phi_R \zeta_R S (\phi_R
\zeta_R))
 - \frac{\delta \lambda_2}{N^2}{\rm tr}(\phi_R \zeta_R) S{\rm tr}(\phi_R
\zeta_R) \nonumber \end{eqnarray}
is invariant under the BRS transformations (\ref{5.6}) together with

\begin{equation}S \psi_R = - \underline{C} \psi_R \; , \quad S \bar{\psi}_R =
\bar{\psi}_R \underline{C} \; . \label{5.6'}\end{equation}

Per say, $Z_3$ and $Z_g$ are independent, however, it emerges below that
the peculiarities of perturbation theory in our approach imply

\begin{equation}\label{5.10}
Z_g \sqrt{Z_3} = 1
\end{equation}
as occurs in the background field gauge \cite{abb}.  With this, dropping the
subscript $R$, the effective lagrangian becomes

\begin{equation}
L^B_{\rm eff} = L_{\rm eff} + \delta L_{\rm eff}
\end{equation}
wherein

\begin{eqnarray}
L_{\rm eff} &=& \bar{\psi} (i \gamma^\mu\underline{D}_\mu - m) \psi -
\frac{1}{2}{\rm tr}
(G^2_{\mu \nu}) + {\rm tr}(\phi D^\dagger_{\mu} D^{\mu} \phi + \{\phi, C\}
D^\dagger_{\mu} D^{\mu} \zeta) \; ,\\
\delta L_{\rm eff} &=& \bar{\psi} ( (Z_2 - 1) i\gamma^\mu \underline{D}_\mu -
(Z_m
- 1) m) \psi - \frac{(Z_3 - 1)}{2}{\rm tr}(G^2_{\mu \nu}) + \nonumber\\
 & &  (\tilde{Z} - 1){\rm tr} (\phi D^\dagger_{\mu} D^{\mu} \phi + \{\phi, C\}
D^\dagger_{\mu} D^{\mu} \zeta) - \nonumber\\
 & &  \frac{\delta \lambda_1}{N}{\rm tr}(\phi^2 + \{\phi, C\} \zeta)^2 -
\frac{\delta
\lambda_2}{N^2} ({\rm tr} (\phi^2 + \{\phi, C\} \zeta))^2\label{5.13}
\end{eqnarray} while the BRS transformations remain those of (\ref{4.19}).

\section{Perturbation theory}

Albeit we have succeeded in our objective of factoring the volume of the gauge
group from $< \underline{O} >$, we would also appear to have painted ourselves
into the proverbial corner in that an inspection of $L_{\rm eff}$ shows that
(i) the term quadratic in the gauge field involves the transverse projection
operator ($g_{\mu \nu} - \Box^{-1} \partial_{\mu} \partial_{\nu}$) so the
corresponding propagator does not exist, and (ii) there is no kinetic term for
the ghosts while $L_{\rm eff}$ is a polynomial higher than quartic in the
fields.  Actually the latter is an unfortunate byproduct of our transformation
from $\zeta^\dagger$ to $C$ and can be avoided by retaining $\zeta^\dagger$ and
$\zeta$ as the ghosts.

In contrast, the problem with the gauge field is profound; to redress this
situation suppose

\begin{equation}\label{6.1}
< \phi > = v I \; .
\end{equation}
Decomposing $\phi$ as

\begin{equation}\label{6.2}
\phi = \phi + v I = \phi_a t_a + v I
\end{equation}
and denoting

\begin{equation}
m_{_G} = gv
\end{equation}
one has

\begin{eqnarray} L_{\rm eff} &=& L + m^2_G{\rm tr} (G_{\mu} G^{\mu}) + {\rm
tr}
(\phi D^\dagger_{\mu} D^{\mu} \phi) + 2 g m_{_G}{\rm tr} (G_{\mu} G^{\mu} \phi)
+ 2v {\rm tr}
(C D^\dagger_{\mu} D^{\mu} \zeta) + \nonumber\\ & &  {\rm tr} (\{\phi, C \}
D^\dagger_{\mu} D^{\mu} \zeta) \; . \end{eqnarray}

Now the gauge field propagator exists as the (unitary gauge) propagator of a
massive vector field

\begin{equation}\label{6.5}
i D^{ab}_{\mu \nu} (k) = \frac{i \delta_{ab}}{k^2 - m^2_G} \left[- g_{\mu \nu}
+ \frac{k_{\mu} k_{\nu}}{m^2_G}\right]
\end{equation}
and we have an identifiable, if unconventional, kinetic term for the ghosts.

Since for a BRS invariant vacuum $0 = < SF > = S < F >$, (\ref{6.1}) requires
$S v = 0$.  Then $L_{\rm eff}$ and the measure remain invariant under the
amended BRS transformations

\begin{eqnarray}\label{6.6} S v &=& 0 \nonumber\\ S \phi &=& - vC - C \phi
\nonumber\\ S C &=& - C C \nonumber\\S \zeta &=& - \phi
- vI - C \zeta\\ S G_{\mu} &=& - \frac{i}{g} [D_{\mu}, C] \nonumber\\ S \psi
&=&
- \underline{C} \psi \; , \quad S \bar{\psi} \; = \; \bar{\psi} \underline{C}
\nonumber \end{eqnarray}

That we may assign a vacuum expectation value to the scalar field while leaving
(a version of) the BRS invariance intact should come as no surprise since a
similar situation exists in supersymmetric theories.  Crucial here is that from
the BRS identity $0 = S {\rm tr} < (v - \phi) \zeta >$ one obtains

\begin{equation}
N v^2 = {\rm tr} < \phi^2 + \{\phi, C \} \zeta > \; ;
\end{equation}
as the right hand side is of order $\hbar$ the tree--level breaking of the
gauge symmetry is ruled out.

Of course this does not deny the possibility of a dynamical mechanism for
generating $v$, one such being that of Coleman and Weinberg \cite{col}:
(\ref{6.2}) injected in (\ref{5.13}) leads to

\begin{eqnarray}\label{6.8}\delta L_{\rm eff} &=& (\tilde{Z} - 1) m^2_G {\rm
tr}(G_{\mu}
G^{\mu}) + (\delta \lambda_1 + \delta \lambda_2) v^4 - \nonumber\\ & & \left(6
\frac{\delta \lambda_1}{N} v^2 + 2 \frac{\delta \lambda_2}{N} v^2 \right){\rm
tr} \phi^2 - \left(4 \frac{\delta \lambda_2}{N} v^2 \right) \frac{1}{N} ({\rm
tr} \phi)^2 - \nonumber\\ & &  \left(2 \frac{\delta \lambda_1}{N} v^2 + 2
\frac{\delta \lambda_2}{N} v^2 \right){\rm tr} (2 vC \zeta) + \dots
\end{eqnarray}

Using $n (= 4 - 2\epsilon)-$dimensional regularization, finiteness of the
one--loop
scalar self energy, Figure 1a, at zero momentum gives the modified minimal
subtraction constants

\begin{equation}\label{6.9} \delta \lambda_1 = \delta \lambda_2 = \frac{3 N^2
g^4}{128
\pi^2} \left(\frac{1}{\epsilon} - \gamma + \ln 4 \pi \right) \end{equation}
which also yield a finite zero momentum ghost self energy, Figure 1b, and a
finite effective potential for $v$

\begin{equation}
V_{\rm eff} (v) = \frac{3 N^2 g^4}{6 4 \pi^2} v^4 \left[\ln \left(\frac{g^2
v^2}{\mu^2}\right) - \frac{5}{6}\right] \; , \end{equation}
$\mu$ being the dimensional regularization scale.  This effective potential has
a minimum for $\ln (m^2_G/\mu^2) = \frac{1}{3}$ which gives the scalar mass
matrix

\begin{equation} (m^2_{\phi})_{ab} = \frac{3 Ng^2}{3 2 \pi^2} m^2_G
(\delta_{ab} + \delta_{a0} \delta_{b0}) \end{equation}
while the ghost mass matrix vanishes.  We observe that the $\delta \lambda_i$
in (\ref{6.9}) coincide with the values obtained by applying Landau gauge field
perturbation theory to $L_{\rm YM} + L_{{\rm aux}}$ .  By matching the vacuum
energy density

\begin{equation}
\epsilon_{{\rm vac}} = - \frac{3 N^2}{128 \pi^2} m^4_G
\end{equation}
to the canonical pure gauge expression \cite{coll}

\begin{equation}\epsilon_{{\rm vac}} = \left< \frac{\beta (g)}{8g} G \cdot G
\right> \label{6.12'}\end{equation}
and using the one--loop $\beta-$function the resulting gluon mass in QCD can be
estimated as

\begin{equation}\label{6.13} m^4_G \approx \frac{4 4 \pi^2}{9N} \left<
\frac{\alpha_s}{\pi} G \cdot G \right>\,. \end{equation} With $\left<
\frac{\alpha_s}{\pi} G \cdot G \right>\approx (330 ~{\rm MeV})^4$ from QCD sum
rules \cite{rein} one finds $m_{_G} \approx 660$ MeV. Of course, since the
corresponding renormalization scale is $\mu=m_{_G}e^{(-1/6)}\approx 560$MeV we
should take this perturbative calculation with a large grain of salt, yet it is
intriguing to note that a relation almost identical to (\ref{6.13}) has been
obtained by Lavelle \cite{lav} from the operator product expansion while a
value $m_{_G}=660\pm80$MeV for the effective gluon mass has been extracted by
Consoli and Field \cite{con} from a study of charmonium decay.

\section{Renormalized gauge field propogator}

Next we face the problems posed by the gauge field propagation (\ref{6.5})
which is order 1 by power counting and through its troublesome longitudinal
part mixes orders in $g$.  Thus, for example, the gauge field contributions to
the gauge field self energy, Figure 2, are

\begin{eqnarray}\label{7.1} &&i \Pi^{\underline{a} \underline{b}}_{_G \mu \nu}
(q)
=\nonumber\\&&\frac{N g^2}{2} \delta_{\underline{a} \underline{b}} \mu^{2
\epsilon}
\int \frac{d^n \ell}{(2 \pi)^n} \left\{ \left(n - 1 - \frac{q^2}{m^2_G}\right)
\left[\frac{(2 \ell + q)_{\mu} (2 \ell + q)_{\nu}} {(\ell^2 - m^2_G) [(\ell +
q)^2- m^2_G]} - \frac{2 g_{\mu \nu}}{\ell^2 - m^2_G}\right] + \right.
\nonumber\\ & & \left.  2 \left(4 - \frac{q^2}{m^2}\right) \frac{(q^2 g_{\mu
\nu} - q_{\mu} q_{\nu})}{(\ell^2 - m^2_G) [(\ell + q)^2- m^2_G]} +
\frac{(\ell_{\mu} q^2 - q_{\mu} \ell \cdot q) (\ell_{\nu} q^2 - q_{\nu} \ell
\cdot q)}{m^4_G (\ell^2 - m^2_G) [(\ell + q)^2 - m^2_G]}\right\} \end{eqnarray}
While transverse, the pole part

\begin{equation} \Pi^{\underline{a} \underline{b}}_{_G \mu \nu} (q) = \frac{N
g^2}{3 2 \pi^2} \frac{\delta_{\underline{a} \underline{b}}}{\epsilon} (q^2
g_{\mu \nu} - q_{\mu} q_{\nu}) \left[7 - \frac{7}{6} \frac{q^2}{m^2_G}
-\frac{1}{12} \left(\frac{q^2}{m^2_G}\right)^2 \right] + \mbox {\rm finite}
\end{equation}
cannot be cancelled by a local counterterm.

All is not lost, however; the gauge field propagation does not belong to the
class $\underline{O}$.  Consider instead the correlator of fermionic
scalar currents

\begin{equation}
{\cal G}(x, y) = < \bar{\psi} (x) \psi (x) \bar{\psi} (y) \psi (y) >
\end{equation}
which does belong to $\underline{O}$.  To order $g^2$ the diagrams
contributing to ${\cal G}$ are given in Figure 3 and it is straightforward to
show that
the contributions due to the longitudinal part of (\ref{6.5}) cancel -- indeed
one may there replace the unitary gauge propagator by

\begin{equation}\label{7.4}
i D_{T_{\mu \nu}}^{\underline{a} \underline{b}} (k) = \frac{i
\delta_{\underline{a} \underline{b}}}{k^2 - m^2_G} \left[- g_{\mu \nu} +
\frac{k_{\mu} k_{\nu}}{k^2}\right] \; .
\end{equation}

Now define

\begin{equation}
{\cal G}_{\infty} (x, y) = \lim_{{x_0 \to + \infty \atop y_0 \to - \infty}}
{\cal G} (x, y) \end{equation}
which clearly also belongs to $\underline{O}$.  Using

\begin{eqnarray}i S_F (x - y) = \int \frac{d^3 p}{(2 \pi)^3} \frac{m}{E}
\sum_{\lambda} [&&\theta (x_0 - y_0) u (p, \lambda) \bar{u} (p, \lambda) e^{- i
p \cdot (x - y)} -\nonumber\\&&\theta (y_0 - x_0) v (p, \lambda) \bar{v} (p,
\lambda) e^{- i p \cdot (x - y)}]\end{eqnarray}
one sees that in ${\cal G}_{\infty} (x, y)$ the fermion lines adjacent to $x$
and $y$ are driven on--shell, i.e., ${\cal G}_{\infty} (x, y)$ is the form of a
convolution with an S-matrix element.  The order $g^2$ graphs contributing to
${\cal G}_{\infty} (x, y)$ are those of Figure 4 for which the replacement
(\ref{7.4}) holds by current conservation.  Indeed the fermion self energy
parts can be exactly cancelled by the fermion mass counterterm, leaving the
first or exchange diagram of Figure 4 with (\ref{7.4}) representing the
effective gauge field propagation.

Of course the above example is somewhat trivial in that to order $g^2\;{\cal G}
(x, y)$ and ${\cal G}_{\infty} (x, y)$ are abelian in character and as is well
known an abelian gauge field mass is harmless.  The important point is that the
analysis extends to order $g^4$ where nonabelian aspects appear.  The order
$g^4$ exchange diagrams contributing to ${\cal G}_{\infty} (x, y)$ are those of
Figure 5.  Note that Figure 5(d) contains Figure 2 as a subgraph.  When the
dangerous longitudinal part of the unitary gauge propagator (\ref{6.5}) is
contracted at a gauge--fermion vertex in Figure 5(a--c) it triggers the simple
Ward identity

\begin{equation}\gamma^\mu k_\mu = S_F^{-1} (k + p) - S^{-1}_F
(p)\end{equation}
to cancel an internal fermion propagator.  The triple gauge vertex has
longitudinal parts which do the same thing; this is the essence of the `pinch
technique' of Cornwall \cite{corn}, and the pinch diagrams corresponding to
Figure
5(a--c) are given in Figure 6 (a--c).  Omitting trivial external factors one
finds for the pinch part,

\begin{eqnarray} 6 ~{\rm (a)} &:& \frac{N}{2} g^2 \delta_{\underline{a}
\underline{b}} \mu^{2\epsilon} \int \frac{d^n \ell}{(2 \pi)^n} \frac{1}{(\ell^2
- m^2_G) [(\ell + q)^2 - m^2_G]} \left\{\frac{\ell_{\mu} \ell_{\nu}}{m^4_G} -
\frac{2 g_{\mu \nu}}{m^2_G} \right\} \nonumber\\ 6 ~{\rm (b)} &:& \frac{N}{2}
g^2 \delta_{\underline{a} \underline{b}} \mu^{2\epsilon} \int \frac{d^n
\ell}{(2 \pi)^n} \left\{\frac{\ell_{\mu} (q^2 \ell_{\alpha} - q_{\alpha} \ell
\cdot q)}{m^4_G (\ell^2 - m^2_G) [(\ell + q)^2 - m^2_G]} \right.- \\ & & \left.
 \frac{\ell_{\mu} (2 \ell + q)_{\alpha} + 2q^2 g_{\mu \alpha}}{m^2_G (\ell^2 -
m^2_G) [(\ell + q)^2 - m^2_G]} + \frac{g_{\mu \alpha}}{m^2_G (\ell^2 - m^2_G)}
\right\} D^{\alpha}_{\nu} (q) \nonumber \end{eqnarray}
while that of Figure 6(c) obtains from 6(b) by $\mu \leftrightarrow \nu$.
Appealing
to current conservation we arrive at the transverse `pinch contribution' to the
gauge field self energy:

\begin{eqnarray}\label{7.7} &&i \Pi^{\underline{a} \underline{b}}_{P \mu \nu}
(q) =\nonumber\\&&\frac{Ng^2}{2} \delta_{\underline{a} \underline{b}}
\mu^{2\epsilon} \int \frac{d^n \ell}{(2 \pi)^n} \left\{ \left(\frac{q^2}{m^2_G}
- 1 \right) \left[\frac{(2 \ell + q)_{\mu} (2 \ell + q)_{\nu}}{(\ell^2 - m
^2_G) [(\ell + q)^2 - m^2_G]} - \frac{2 g_{\mu \nu}}{\ell^2 - m^2_G}\right]
+\right.\nonumber\\&&2 \left(\frac{q^2}{m^2_G} -
\frac{m^2_G}{q^2}\right)\frac{(q^2 g_{\mu \nu} - q_{\mu} q_{\nu})}{(\ell^2 -
m^2_G) [(\ell + q)^2 - m^2_G]} +\nonumber\\ &&\left.\left[\frac{1}{(q^2)^2} -
\frac{1}{m^4_G}\right] \frac{(\ell_{\mu} q^2 - q_{\mu} \ell \cdot q)(\ell_{\nu}
q^2 - q_{\nu} \ell \cdot q)}{(\ell^2 - m^2_G) [(\ell + q)^2 - m^2_G]} \right\}
\end{eqnarray}

The sum of (\ref{7.1}) and (\ref{7.7}) is

\begin{eqnarray}\label{7.8} &&i \Pi^{\underline{a} \underline{b}}_{G + P \mu
\nu} (q)
=\nonumber\\&&\frac{Ng^2}{2} \delta_{\underline{a} \underline{b}}
\mu^{2\epsilon} \int
\frac{d^n \ell}{(2 \pi)^n} \left\{ (n - 2) \left[\frac{(2 \ell + q)_{\mu} (2
\ell + q)_{\nu}}{(\ell^2 - m^2_G) [(\ell + q)^2 - m^ 2_G]} - \frac{2 g_{\mu
\nu}}{\ell^2 - m^2_G}\right] +\right.\nonumber\\&&\left. 2 \left(4 -
\frac{m^2_G}{q^2}\right)
\frac{(q^2 g_{\mu \nu} - q_{\mu} q_{\nu})}{(\ell^2 -
m^2_G) [(\ell + q)^2 - m^2_G]} + \frac{(\ell_{\mu} q^2 - q_{\mu} \ell \cdot
q)(\ell_{\nu} q^2 - q_{\nu} \ell \cdot q)}{(q^2)^2 (\ell^2 - m^2_G) [(\ell +
q)^2 - m^2_G]} \right\} \end{eqnarray}

Still to be included are the ghost and scalar contributions, Figure 7.  The
ghost part of the gauge field self energy is wholly transverse

\begin{equation}\label{7.9} i \Pi^{\underline{a} \underline{b}}_{\zeta \mu \nu}
(q) = -\frac{Ng^2}{2} \delta_{\underline{a} \underline{b}} \mu^{2\epsilon} \int
\frac{d^n
\ell}{(2 \pi)^n} \frac{(2 \ell + q)_{\mu} (2 \ell + q)_{\nu}}{\ell^2 (\ell +
q)^2} \end{equation}
but the scalar part is not

\begin{equation}
i \Pi^{\underline{a} \underline{b}}_{S \mu \nu} (q) = Ng^2
\delta_{\underline{a} \underline{b}} \mu^{2\epsilon} \int \frac{d^n \ell}{(2
\pi)^n}
\left\{\frac{\ell_{\mu} \ell_{\nu} - m^2_G g_{\mu \nu}}{(\ell + q)^2 (\ell^2 -
m^2_G)} - \frac{1}{8} \frac{(2 \ell + q
)_{\mu} (2 \ell + q)_{\nu}}{\ell^2 (\ell + q)^2}\right\}
\end{equation}

On the other hand owing to current conservation at the gauge--fermion vertex we
only need the transverse projection

\begin{equation}\begin{array}{rcl} i\Pi^{\underline{a} \underline{b}}_{ST \mu
\nu} (q) &=& Ng^2 \delta_{\underline{a} \underline{b}} \mu^{2\epsilon}
\displaystyle
\int \displaystyle \frac{d^n \ell}{(2 \pi)^n} \left\{- \displaystyle
\frac{m^2_G}{q^2} \displaystyle \frac{(q^2 g_{\mu \nu} - q_{\mu} q
_{\nu})}{(\ell + q)^2 (\ell^2 - m^2_G)} \right.\\[5mm] & & \displaystyle \left.
+ \frac{(\ell_{\mu} q^2 - q_{\mu} \ell \cdot q) (\ell_{\nu} q^2 - q_{\nu} \ell
\cdot q)}{(q^2)^2 (\ell + q)^2 (\ell^2 - m^2_G)} + \frac{1}{8} \frac{(2 \ell +
q)_{\mu} (2 \ell + q)_{\nu}}{\ell^2 (\ell + q)^2}\right\} \; . \end{array}
\label{7.10'}\end{equation}

The sum of (\ref{7.8}), (\ref{7.9}) and (\ref{7.10'}) gives the effective gauge
field self--energy

\begin{eqnarray}\label{7.11}
\Pi^{\underline{a} \underline{b}}_{\mu \nu} (q) &=& \Pi^{\underline{a}
\underline{b}}_{G + P \mu \nu} (q) + \Pi^{\underline{a} \underline{b}}_{\zeta
\mu \nu} (q) + \Pi^{\underline{a} \underline{b}}_{ST \mu \nu} (q) \nonumber\\
&=& \delta_{\underline{a} \underline{b}} (q^2 g_{\mu \nu} - q_{\mu} q_{\nu})
\Pi (q^2) \end{eqnarray}
and one finds

\begin{equation} \Pi (q^2) = \frac{Ng^2}{16 \pi^2} \left(\frac{11}{3} -
\frac{3}{2} \frac{m^2_G}{q^2}\right) \left(\frac{1}{\epsilon} - \gamma + \ln
4 \pi \right) + \mbox {\rm finite} \; . \end{equation}

Finally there are the field and gauge mass counterterms implied by (\ref{5.13})
and (\ref{6.8}) respectively -- for the latter we again need the transverse
projection so

\begin{equation}
\delta \Pi (q^2) = (\tilde{Z} - 1) \frac{m^2_G}{q^2} - (Z_3 - 1)
\end{equation}

Thus we see that

\begin{equation}
\Pi_R (q^2) = \Pi (q^2) + \delta\Pi (q^2)
\end{equation}
is rendered finite for

\begin{eqnarray} \tilde{Z} &=& 1 + \frac{3 N g^2}{32 \pi^2}
\left(\frac{1}{\epsilon} - \gamma + \ln 4 \pi \right) \; ,\\ Z_3 &=& 1
+ \frac{11 N g^2}{48 \pi^2} \left(\frac{1}{\epsilon} - \gamma + \ln \pi
\right)\label{7.16} \; , \end{eqnarray}
and so is the effective one--loop renormalized gauge field propagator

\begin{equation}
i D_{\perp \mu \nu}^{\underline{a} \underline{b}} (k) = \frac{i
\delta_{\underline{a} \underline{b}}}{k^2 [1 + \pi_R (k^2)] - m^2_G} \left[-
g_{\mu \nu} + \frac{k_{\mu} k_{\nu}}{k^2}\right] \; ;
\end{equation}
this is our principle perturbative result.  We note that as for $\delta
\lambda_1$ and $\delta \lambda_2$, $\tilde{Z}$ agrees with the value obtained
from perturbative Landau gauge fixing applied to $L_{\rm YM} + L_{{\rm aux}}$ .

Further, as promised, via (\ref{5.10}) $Z_3$ of (\ref{7.16}) contains the full
information regarding the one--loop gauge coupling renormalization.  This
feature may be traced to the role of the longitudinal pieces of the triple
gauge vertex in the pinch program; an inspection of the calculation in
\cite{corn}
shows that in gauge fixed perturbation theory the gauge invariant proper
vertices laboriously constructed from the pinch technique coincide with what is
obtained through the background field method of Abbott \cite{abb}.

Indeed, from a strict perturbative viewpoint -- i.e. leaving aside possible
dynamical mechanisms for generating $v \neq 0$ -- one could proceed from the
minimal identity of section 2, introducing $v$ as in (\ref{6.2}) through
(\ref{6.6}) with
the meaning of a `gauge parameter' to define the intermediate steps with the
limit $v \to 0$ (after pinching to remove the pieces singular as $v \to 0$)
understood.  Taking $m_G \to 0$ in (\ref{7.11})

\begin{equation}
\Pi (q^2) \longrightarrow  - i \frac{Ng^2}{2} \left[8 -
\left(\frac{n - 2}{n - 1}\right) \right] \mu^{2\epsilon} \int \frac{d^n
\ell}{(2 \pi)^n} \frac{1}{\ell^2 (\ell + q)^2} \end{equation}
and

\begin{equation} \Pi_R (q^2) \longrightarrow \frac{Ng^2}{16
\pi^2} \left[- \frac{11}{3} \ln \left(\frac{- q^2}{\mu^2}\right) +
\frac{67}{9}\right] \end{equation}
which are canonical to the background field gauge.

Another perspective is gained by observing that because the amended BRS
symmetry (\ref{6.6}) is non-intrinsic $S$ and $\frac{\partial}{\partial v}$ do
not commute, but rather

\begin{equation}
\frac{\partial}{\partial v} SW=S\frac{\partial}{\partial v}W-2ig{\rm
tr}(G^\mu\partial_\mu\phi)
\end{equation}
there follows, integrating by parts,

\begin{eqnarray}
\frac{\partial}{\partial v} \ln Z&=&-<g\int d^4x\phi_a\partial\cdot
G^a>\label{7.21}\,,\\ \frac{\partial}{\partial
v}<\underline{O}>&=&<\underline{O}><g\int d^4x\phi_a\partial\cdot
G^a>- <\underline{O}\,g\int d^4x\phi_a\partial\cdot G^a>
\end{eqnarray}
and in the perturbative regime where we can apply ordinary gauge fixing in the
form $\partial\cdot G=0$ both $\partial\ln Z/\partial v$ and
$\partial\langle O\rangle/\partial v$ vanish.  Moreover, even when mass is
dynamically
generated through the Coleman-Weinberg mechanism as discussed in the preceding
section $\partial\ln Z/\partial v$ must vanish since $\epsilon_{\rm vac}$ is an
extremum; one readily verifies that the diagrams of figure 8 give a null result
for the right hand side of (\ref{7.21})

\section{Discussion and conclusions}

We have shown how a SU(N) gauge theory can be quantized in the continuum
without fixing the gauge.  In this way the Gribov problem is
circumvented.  The underlying BRS symmetry has been identified and the
associated Slavnov-Taylor identities can be derived in the usual way.  We
used this BRS symmetry to perform the perturbative renormalization of the
effective theory and showed that in the perturbative regime the procedure is
equivalent to Landua gauge fixing.  All the usual perturbative results
can therefore be recovered.

\section{Acknowledgements}

This work was supported by a grant from the Foundation of Research Development.

\newpage
\appendix
\section{}

Herein we collect some notation and useful relations for the groups U(N) and
SU(N)$\in$U(N).  In the fundamental representation the U(N) generators
are denoted

\begin{equation}t_0 = 1/\sqrt{2N} \quad \mbox{\rm ,} \quad t_{\underline{a}} =
\lambda_{\underline{a}}/2 \label{A.1}\end{equation}
where (non) underlined indices run from (0) 1 to $N^2 - 1$; the
$\lambda_{\underline{a}}$'s are the SU(N) generalizations of the
Gell--Mann
matrices, ${\rm tr} (\lambda_{\underline{a}} \lambda_{\underline{b}}) = 2
\delta_{\underline{a} \underline{b}}$, so the $t_a$'s are normalized to

\begin{equation}{\rm tr} (t_a t_b) = \frac{1}{2} \delta_{ab}
\label{A.2}\end{equation}
and generate the U(N) algebra

\begin{equation}\begin{array}{rclcrcl} \displaystyle [t_a, \, t_b] &=& i
F_{abc} t_c &,& F_{ab0} &=& 0\\[2mm] \displaystyle \{t_a, \, t_b\} &=& D_{abc}
t_c &,& D_{ab0} &=& \displaystyle \sqrt{\frac{2}{N}} \delta_{ab} \end{array}
\label{A.3}\end{equation}
with ($F_{abc}$) $D_{abc}$ the totally (anti) symmetric structure constants.
Clearly the $t_{\underline{a}}$ generate the SU(N) subalgebra.

\begin{equation}\begin{array}{rclcrcl} \displaystyle [t_{\underline{a}}, \,
t_{\underline{b}}] &=& i f_{\underline{a} \underline{b} \underline{c}}
t_{\underline{c}} &,& f_{\underline{a} \underline{b} \underline{c}} &=&
F_{\underline{a} \underline{b} \underline{c}}\\[3mm] \displaystyle
\{t_{\underline{a}}, \, t_{\underline{b}}\} &=& \displaystyle
\frac{\delta_{ab}}{N} + d_{\underline{a} \underline{b} \underline{c}}
t_{\underline{c}} &,& d_{\underline{a} \underline{b} \underline{c}}
&=& D_{\underline{a} \underline{b} \underline{c}} \end{array}
\label{A.4}\end{equation}

By completeness

\begin{equation}(t_a)_{ij} (t_a)_{k \ell} = \frac{1}{2} \delta_{ij} \delta_{k
\ell} \; , \quad i, j = 1, \dots , N \; ; \label{A.5}\end{equation}
hence

\begin{equation}t_a t_a = \frac{N}{2} \quad \mbox{\rm ,} \quad t_a t_b t_a =
\frac{1}{2} \sqrt{\frac{N}{2}} \delta_{b0} \label{A.6}\end{equation}
while for the SU(N) subgroup

\begin{equation}\begin{array}{rclcrcl} \displaystyle t_{\underline{a}}
t_{\underline{a}} &=& C_F, && C_F &=& \displaystyle \frac{N^2 - 1}{2N}\\[3mm]
\displaystyle t_{\underline{a}} t_{\underline{b}} t_{\underline{a}} &=& (C_F -
C_A/2) t_{\underline{b}}, && C_A &=& N \end{array} \label{A.7}\end{equation}
Then

\begin{eqnarray}
F_{acd} F_{bcd} &=& - 2{\rm tr}([t_a, t_c] [t_b, t_c])\nonumber\\
&=& 2{\rm tr}(\{t_a, t_b\} t_c t_c - 2 t_a t_c t_b t_c)\nonumber\\
&=& N (\delta_{ab} - \delta_{a0} \delta_{b0})
\equiv \; N \delta_{\underline{a} \underline{b}}\label{A.8}
\end{eqnarray}
and similarly

\begin{equation}D_{acd} D_{bcd} = 2 {\rm tr}(\{t_a, t_c\} \{t_b,
t_c\})= N (\delta_{ab} + \delta_{a0} \delta_{b0})\label{A.9} \end{equation}

Now, define the $N^2 \times N^2$ matrices $T_a$ as

\begin{equation}(T_a)_{bc} \equiv \frac{1}{2} (D_{abc} - i F_{abc})
\label{A.10}\end{equation}
so from (\ref{A.3})

\begin{equation}t_a t_b = (T_b)_{ac} t_c \; . \label{A.11}\end{equation}

There follows

\begin{equation}{\rm tr}(t_a t_b t_c) = \frac{1}{2} (T_b)_{ac}
\label{A.12}\end{equation}
and

\begin{eqnarray}
{\rm tr}(t_a t_b t_c t_d) &=& \frac{1}{2} (T_b)_{ac} (T_c)_{ed}\nonumber
\\[2mm] &=& \frac{1}{2} (T_a)_{de} (T_b)_{ec}\nonumber\\
&=&  i F_{bce} \frac{1}{2} (T_e)_{ad} +
\frac{1}{2} (T_c)_{ae} (T_b)_{ed} \label{A.13}
\end{eqnarray}
from which

\begin{equation}\begin{array}{rcl}
[T_a, T_b] &=& i F_{abc} T_c \; , \\[1mm]
{}~[T_a, - T^{\ast}_b] &=& 0 \; , \\[1mm]
{}~[- T^{\ast}_a, - T^{\ast}_b] &=& i F_{abc} (- T^{\ast}_c) \; ,
\end{array} \label{A.14}\end{equation}
i.e., the $T_a$ and $- T^{\ast}_a$ generate the chiral algebra U(N)$_{\rm L}
\otimes$ U(N)$_{\rm R}$.  They are normalized by

\begin{equation}{\rm tr}(T_a T_b) = \frac{N}{2} \delta_{ab} ={\rm
tr}(T^{\ast}_a T^{\ast}_b) \; , \label{A.15}\end{equation}
as a consequence of (\ref{A.8}) and (\ref{A.9}).  The diagonal subalgebra
generated by

\begin{equation}(T_a - T^{\ast}_a)_{bc} = - i F_{abc}
\label{A.16}\end{equation}
is isomorphic to the adjoint representation of SU(N).

\newpage

\section{}

Consider the change of variables (\ref{2.18});  we define

\begin{eqnarray}\ell_a (\pi) &\equiv& U^\dagger (\pi) \frac{\partial}{\partial
\pi_a} U (\pi)\nonumber\\
&=& \int^1_0 d \lambda U^\dagger (\lambda \pi) i t_a U
(\lambda \pi) \label{B.1}\end{eqnarray}
and note by repeated use of (\ref{A.11})

\begin{equation}t_a U (\pi) = U_{ab} (\pi) t_b \label{B.2}\end{equation}
where $U (\pi)$ is the $N^2 \times N^2$ matrix

\begin{equation}U (\pi) = \mbox {\rm exp} ~(i \pi_a T_a) \; .
\label{B.3}\end{equation}

Then

\begin{equation}\begin{array}{rcl} \ell_a (\pi) &=& L_{ab} (\pi) i t_b \; = \;
i t_b L_{ba} (- \pi) \; ,\\[3mm] L (\pi) &=& \displaystyle \int^1_0 d \lambda
\, \mbox {\rm exp} ~(i \lambda\pi_a (T_a - T_a^{\ast})) \end{array}
\label{B.4}\end{equation} and since

\begin{eqnarray} {\rm tr}(\partial_{\mu} U^\dagger (\pi) \partial^{\mu} U
(\pi)) &=& -{\rm tr} (\ell_a \ell_b) \partial_{\mu} \pi_a \partial^{\mu}
\pi_b\nonumber\\ &=& \frac{1}{2} (L (\pi) L (- \pi))_{ab} \partial_{\mu} \pi_a
\partial^{\mu} \pi_b\nonumber\\
&\equiv& \frac{1}{2} g_{ab} (\pi) \partial_{\mu} \pi_a \partial^{\mu} \pi_b
\label{B.5}\end{eqnarray}
the invariant measure on the U(N) group space is

\begin{equation}dU = \sqrt{\mbox {\rm det} ~(g (\pi))}\, \Pi_a d \pi_a =
\mbox {\rm det} ~(L (\pi)) \Pi_a d \pi_a \; . \label{B.6}\end{equation}

Now

\begin{equation}\frac{\partial}{\partial \pi_a} U (\pi) = U (\pi) \ell_a (\pi)
\; , \quad \frac{\partial}{\partial \pi_a} U^\dagger (\pi) = - \ell_a (\pi)
U^\dagger (\pi) \label{B.7}\end{equation}
so

\begin{equation}\begin{array}{rclcl} \partial \Phi_a / \partial \pi_b &=& 2{\rm
tr} ( - t_a U (\pi) \ell_b (\pi) \phi) &=& (- U (\pi) i \phi_e T^{\ast}_e L (-
\pi))_{ab} \; ,\\[2mm] \partial \Phi_a / \partial \phi_b &=& 2{\rm tr} ( - t_a
U (\pi) t_b) &=& - U_{ab} (\pi) \; ,\\[2mm] \partial \Phi^\dagger_a / \partial
\pi_b &=& 2{\rm tr}(\phi \ell_b (\pi) U^\dagger (\pi) t_a) &=& (U^{\ast} (\pi)
i \phi_e T_e L (- \pi))_{ab}\\[2mm] \partial \Phi^\dagger_a / \partial \phi_b
&=& 2{\rm tr}( - t_b U^\dagger (\pi) t_a) &=& - U^{\ast}_{ab} (\pi) \end{array}
\label{B.8}\end{equation}
and the jacobian factorizes:

\begin{eqnarray} \Pi_a \frac{d \Phi_a d \Phi^\dagger_a}{2 \pi i} &=& \mbox {\rm
det} \left(\begin{array}{ccc} & | & \\ U (\pi) & | & 0\\ - - - - -
\hspace*{-3.5mm} & | & \hspace*{-3.5mm} - - - -\\ 0 & | & U^{\ast} (\pi)
\end{array} \right) \, \mbox {\rm det} ~\left(\begin{array}{ccc} & | & \\ -
\phi_e T^{\ast}_e L (- \pi) & | & - I\\ - - - - - - - \hspace*{-4mm} & | &
\hspace*{-4mm} - - -\\ \phi_e T_e L (- \pi) & | & - I \end{array} \right)
{}~\Pi_a \frac{d \phi_a}{2 \pi} d \pi_a\nonumber\\ &=& \mbox {\rm det} ~(U
(\pi)
U^T (\pi)) \, \mbox {\rm det} ~(\phi_e T^{\ast}_e + \phi_e T_e) \, \mbox {\rm
det} ~(L (\pi)) \Pi_a \frac{d \phi_a}{2 \pi} d \pi_a\nonumber\\
&=& dU \, \mbox {\rm det} ~(J (\phi)) \Pi_a \frac{d \phi_a}{2 \pi}
 \label{B.9}\end{eqnarray}
with

\begin{equation}J_{ab} (\phi) = D_{abc} \phi_c \label{B.10}\end{equation}

\newpage

\narrowtext

\begin{figure}
\caption{One-loop diagrams contributing to (a) the scalar and (b) the ghost
self-energy; wavey lines denote the guage field.}
\label{fig1}

\end{figure}

\begin{figure}
\caption{Gauge field contributions to the gauge field self-energy.}
\label{fig2}

\end{figure}

\begin{figure}
\caption{Order $g^2$ (two-loop) contributions to ${\cal G}(x,y)$.}
\label{fig3}

\end{figure}

\begin{figure}
\caption{Order $g^2$ graphs contributing to ${\cal G}_{\infty}(x,y)$.}
\label{fig4}

\end{figure}

\begin{figure}
\caption{Order $g^4$ graphs contributing to the exchange part of ${\cal
G}_{\infty}(x,y)$.} \label{fig5}

\end{figure}

\begin{figure}
\caption{Pinch parts of the exchange diagrams contributing to ${\cal
G}_{\infty}(x,y)$ in order $g^4$.} \label{fig6}

\end{figure}

\begin{figure}
\caption{(a) Ghost and (b) scalar contributions to the exchange part of ${\cal
G}_{\infty}(x,y)$ in order $g^4$.} \label{fig7}

\end{figure}

\begin{figure}
\caption{Two-loop diagrams contributing to $\partial\ln Z/\partial v$.  The
square denotes the operator insertion of (126).} \label{fig8}

\end{figure}

\end{document}